\documentclass{webofc}
\usepackage[colorinlistoftodos]{todonotes}
\usepackage{amsmath}
\usepackage{booktabs}
\usepackage{wrapfig}
\usepackage[rightcaption]{sidecap} 
\usepackage[normalem]{ulem}
\usepackage[varg]{txfonts}
\usepackage{caption}
\usepackage{braket}
\usepackage{subcaption}

\usepackage{hyperref}
\hypersetup{
    colorlinks,
    citecolor=black,
    linkcolor=black,
    urlcolor=black,
    pdfencoding=auto,
    pdfstartpage=1,
}

\begin{document}

\title{Higgs analysis with quantum classifiers}

\author{\firstname{Vasilis} \lastname{Belis}\inst{1}\fnsep\thanks{\email{vasileios.belis@cern.ch}} \and
        \firstname{Samuel} \lastname{González-Castillo}\inst{2} \and
        \firstname{Christina} \lastname{Reissel}\inst{1} \and
        \firstname{Sofia}
        \lastname{Vallecorsa}\inst{3} \and
        \firstname{Elías}
        \lastname{F. Combarro} \inst{4}\and
        \firstname{Günther}
        \lastname{Dissertori}\inst{1}\and
        \firstname{Florentin}
        \lastname{Reiter} \inst{5}
}

\institute{Institute of Particle Physics and Astrophysics, ETH Zürich, Zürich, Switzerland
\and
Faculty of Sciences, University of Oviedo, Oviedo, Spain
\and
CERN, 1, Esplanade des Particules, Geneva, CH 1211
\and
Department of Computer Science, University of Oviedo, Oviedo, Spain
\and
Institute for Quantum Electronics, ETH Zürich, Zürich, Switzerland
}

\abstract{%
  We have developed two quantum classifier models for the $t\bar{t}H(b\bar{b})$ classification problem, both of which fall into the category of hybrid quantum-classical algorithms for Noisy Intermediate Scale Quantum devices (NISQ). Our results, along with other studies, serve as a proof of concept that Quantum Machine Learning (QML) methods can have similar or better performance, in specific cases of low number of training samples, with respect to conventional machine learning (ML) methods even with a limited number of qubits available in current hardware. To utilise algorithms with a low number of qubits --- to accommodate for limitations in both simulation hardware and real quantum hardware --- we investigated different feature reduction methods. Their impact on the performance of both the classical and quantum models was assessed. We addressed different implementations of two QML models, representative of the two main approaches to supervised quantum machine learning today: a Quantum Support Vector Machine (QSVM), a kernel-based method, and a Variational Quantum Circuit (VQC), a variational approach.

}

\maketitle
\section{Introduction}
\label{intro}

Identifying the Higgs boson production in association with top quark - antiquark pairs in which the Higgs decays into a pair of bottom quark-antiquark allows studying the Yukawa coupling of the Higgs boson in a purely fermionic process. Precise measurements of the $t\bar{t}H$ production processes with the LHC experiments at CERN are crucial tests of our fundamental understanding of the universe and serve as probes for potential undiscovered physics, as the Higgs boson - top quark Yukawa coupling directly carries information about the scale of new physics \cite{Bezrukov_2015}.

\begin{figure}[h]
    \centering
    \sidecaption
    \includegraphics[width=0.45\textwidth]{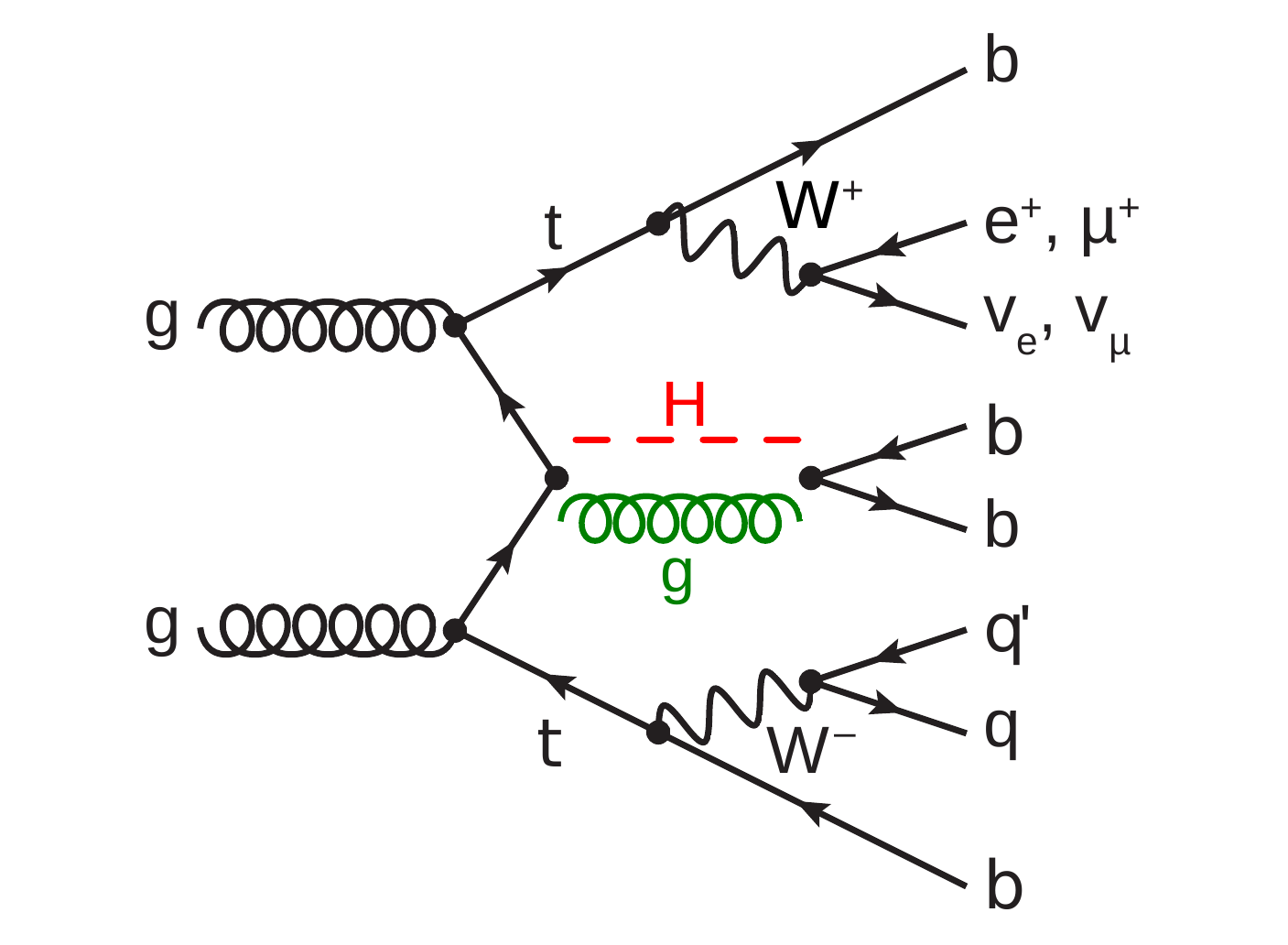}
    \caption{Example of Leading Order (LO) Feynman diagram of the signal process in red and the dominant background process in green. The Higgs Boson is produced in association with $t\bar{t}$ via gluon fusion and it decays to $b\bar{b}$. The channel is semi-leptonic as only one of the $W$ bosons decays into leptons.}
    \label{fig:feynman}
\end{figure}

The considered signal and background processes, in red and green respectively, are depicted in Fig.\ \ref{fig:feynman}. Distinguishing signal over background is extremely challenging since in both cases the final state is the same. The complex final state of the $t\bar{t}H(b\bar{b})$ process comes with a large number of jets, but allows studying the purely fermionic Higgs production and decay. The semi-leptonic channel is addressed in order to suppress the QCD background. Moreover, using the decay of the Higgs boson into a pair of bottom quarks balances the very low production cross section leading to larger number of events to be observed. 

Classification of $t\bar{t}H(b\bar{b})$ versus the dominant $t\bar{t}b\bar{b}$ background is typically addressed with Multivariate Analysis techniques (MVA). Current physics analyses are using analytic methods like the Matrix-Element Method~\cite{Artoisenet2013}, as well as Boosted Decision Trees (BDTs) and Neural Networks (NN) to tackle the discrimination of signal and the overwhelming background~\cite{CMS-PAS-HIG-18-030,ATLAS-CONF-2020-058}.

 We developed two quantum classifier models for the $t\bar{t}H$ classification problem, both of which fall into the category of hybrid quantum-classical algorithms for Noisy Intermediate Scale Quantum devices (NISQ): a Quantum Support Vector Machine (QSVM), a kernel-based method, and a Variational Quantum Circuit (VQC), a variational approach.
 In order to reduce the problem dimensionality and ease the classical input feature encoding in quantum states, we extract a compressed representation using a latent space. The performance of the classical methods, mentioned above, is also measured on this reduced data representation. 

This paper is organised as follows: the data set and the pre-processing step are described in section \ref{sec:dataset}; a description of the two quantum algorithms follows in section \ref{sec:qclas}; section \ref{sec:res} contains an analysis of the results including a detailed comparison to the classical benchmarks. Finally section \ref{sec:conc} summarizes our findings and our future plans.

\section{The data set and pre-processing step} \label{sec:dataset}
The Monte Carlo (MC) data samples used in the study are generated using Powheg v.2~\cite{Nason_2004,Frixione_2007,Alioli2010} for hard scattering, Pythia 8 \cite{pythia} for parton shower simulations and Delphes v.3.4.1 \cite{delphes} for the detector response simulation. Delphes is configured with the CMS detector topology and Run II settings.

\subsection{Pre-selection cuts}
In the pre-selection step, we require the electrons and muons to pass selection criteria of the transverse momentum $p_T$, pseudorapidity $\eta$ and isolation with respect to jets. Namely, the object selection cuts are: $p_T>30$ GeV, $|\eta|<2.1$ and iso $>0.1$ for the electrons, $p_T>26$ GeV, $|\eta|<2.1$ and iso $>0.1$ for the muons and $p_T>30$ GeV, $|\eta|<2.4$ for the jets. After this initial object selection, we further consider events with at least one lepton and at least 4 jets from which at least two are b-tagged. Thus, the following event selection criteria are used: $n^\text{jet}\geq 4$,  $n^\text{b-tag}\geq 2$ and $n^\text{leptons}=1$. From the leading-order description of the process depicted in Fig. \ref{fig:feynman} we identify that the nominal expectation consists of 4 b-tagged jets and 2 jets of any flavour, hence 6 jets in total. For our analysis, we keep the 7 most energetic jets of each event, allowing one extra jet to take into account initial or final state radiation.  

\subsection{Auto-Encoders for feature reduction}\label{sec:autoencoder}

Feature selection, i.e. choosing a set of physical observables to be used as an input for the classifier model, is of particular importance in every HEP analysis. Features are chosen according to their discriminative power, meaning that they enable the classifier to best discriminate the signal against the background events. 
Realistic quantum methods that can be used in Noisy Intermediate-Size Quantum (NISQ) devices are restricted to a relatively low dimensional feature space which also limits the information given to the classifiers. 
Most studies proposing QML methods in HEP  involve selecting only a subset of available observables \cite{blance2020quantum,terashi2021} or deploy a feature reduction method, such as Principle Component Analysis (PCA) \cite{wu2020application}. 

We implemented two Auto-Encoder neural networks~\cite{DLbook} for the $t\bar{t}H(b\bar{b})$ analysis to reduce the dimension of the feature space from  67 --- representing the kinematics of jets, leptons and missing transverse momentum --- to 16 and 8 dimensions in the latent space, respectively. The Auto-Encoder inherently takes into account non-linear correlations between the observables; thus, in principle,  higher order correlations between the features are not lost like in the case of PCA. Due to this property, we expect that as much information as possible is maintained from the original features.

For the Auto-Encoder training and testing, the MC dataset is split into training (80\%), validation (10\%) and test (10\%) data sets, with sizes $N^\text{train} = 1.15\times 10^6$ and $N^\text{validation}=N^\text{test} = 1.44\times 10^5$. The hyper-parameters of the two Auto-Encoder models, one implemented using PyTorch and the other using TensorFlow, are presented in Table \ref{table:ae}. After each training epoch, the loss is computed on the validation set. The chosen model is the one that has the lowest Mean Squared Error (MSE) on the validation data set.
The final layer and the latent layer, i.e.\ the layer representing the latent space, of both Auto-Encoders uses a Sigmoid activation function. Thus guaranteeing that the latent space features and the output of the decoder are bounded between $0$ and $1$.

\begin{figure}[h]
\begin{subfigure}{0.47\textwidth}
    \centering
    \includegraphics[width=\textwidth]{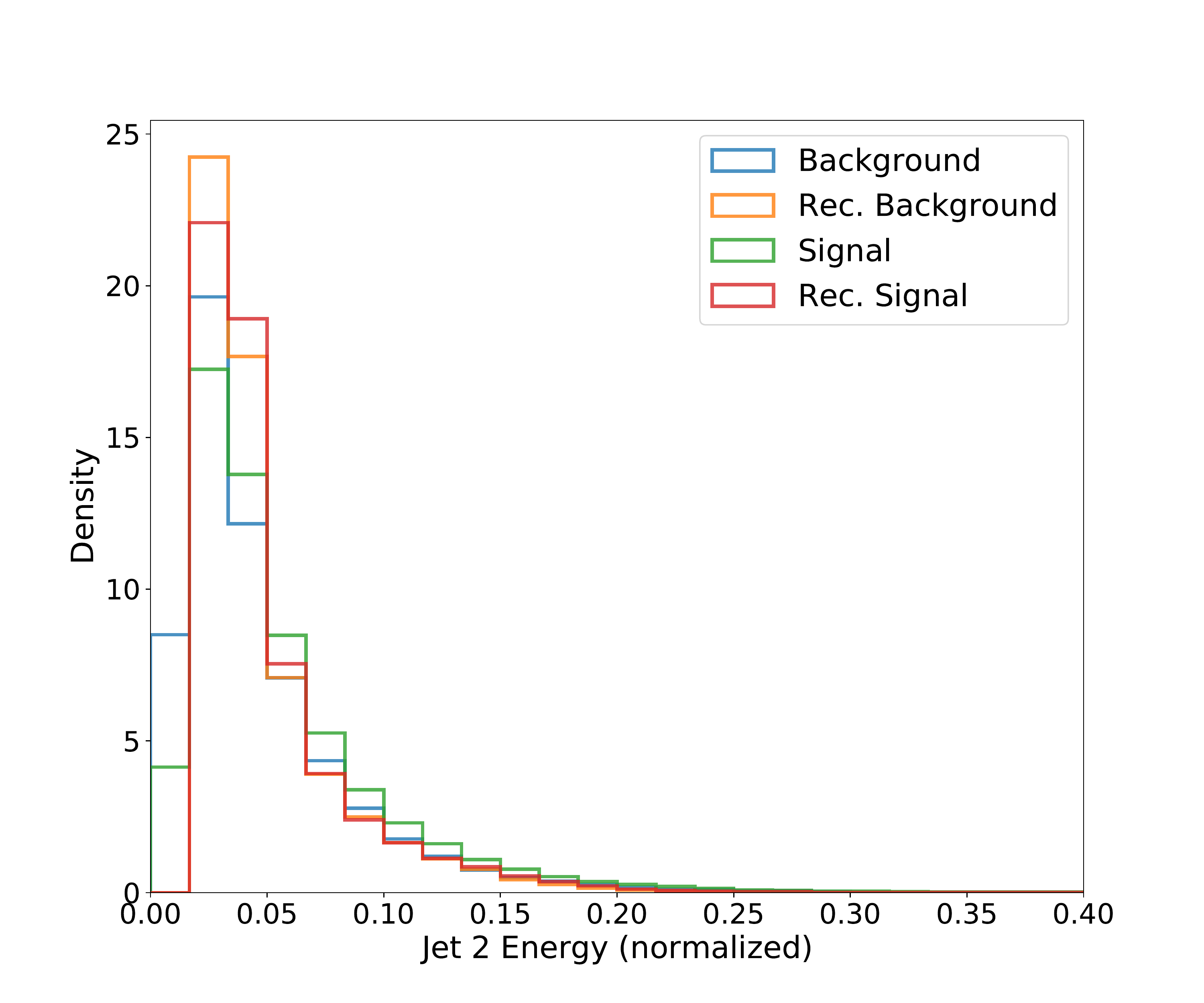}
    \label{fig:pt_reco}
    \caption{PyTorch Auto-Encoder}
\end{subfigure}
\hspace{.06\textwidth}
\begin{subfigure}{0.47\textwidth}
    \centering
    \includegraphics[width=\textwidth]{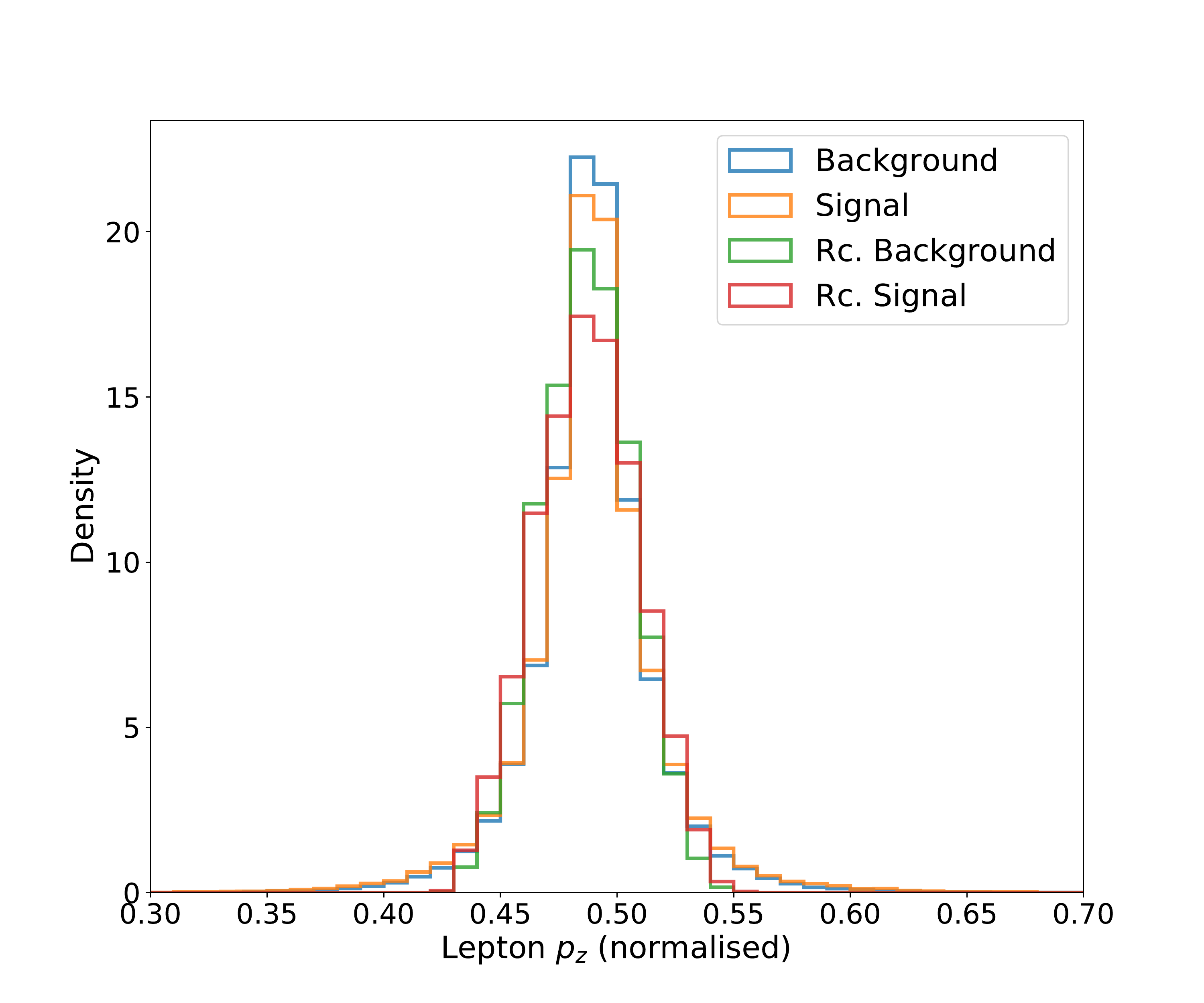}
    \label{fig:tf_reco}
    \caption{TensorFlow Auto-Encoder}
\end{subfigure}
\caption{Histogram representing the original and reconstructed (Rec./Rc.) data for one example variable using the two Auto-Encoders.}
\label{fig-ae}
\end{figure}

\begin{SCtable}[][h]

\begin{tabular}{l|c|c|}
\cline{2-3}
                                            & \multicolumn{1}{l|}{PyTorch AE} & \multicolumn{1}{l|}{TensorFlow AE} \\ \hline
\multicolumn{1}{|l|}{Layer Type}            & \multicolumn{2}{c|}{Dense}                                           \\ \hline
\multicolumn{1}{|l|}{Encoder hidden layers} & 6                               & 7                                 \\ \hline
\multicolumn{1}{|l|}{Latent space dim.}     & 16                              & 8                                  \\ \hline
\multicolumn{1}{|l|}{Loss}                  & \multicolumn{2}{c|}{Mean Square Error (MSE)}                         \\ \hline
\multicolumn{1}{|l|}{Optimizer}             & \multicolumn{2}{c|}{Adam}                                            \\ \hline
\multicolumn{1}{|l|}{Learning Rate}         & $2\times 10^{-3}$                           & $\sqrt{3} \times 10^{-3}$                                 \\ \hline
\multicolumn{1}{|l|}{Batch size}            & 128                             & 93                                 \\ \hline
\multicolumn{1}{|l|}{Number of epochs}      & 80                              & 30                                 \\ \hline
\end{tabular}
\hspace{0.5cm}
\caption{Autoencoder (AE) hyperparameters. The decoder part of the network is symmetric with respect to the encoder in terms of layers and node numbers. The hyperparameter tuning is done by semi-grid search.}
\label{table:ae}
\end{SCtable}


The MSE obtained on the test sample with the PyTorch  Auto-Encoder is $6.4\times10^{-4}$, while that of the TensorFlow Auto-Encoder is $5.07\times10^{-3}$. Figure~\ref{fig-ae} shows histograms comparing an example distribution, from the original data, and the corresponding results, obtained after reconstruction (encoding and decoding) with the two Auto-Encoders.

The latent space features serve as the input of the quantum circuits used in the classifier models, the QSVM and the VQC, and are embedded, or equivalently encoded, via different feature maps. We use SVM and other classical models for benchmarking the quantum models with respect to ROC curves, and the corresponding AUC, measured on the test data sets.

\begin{figure}[h]
    \centering
    \sidecaption
    \includegraphics[width=0.5\textwidth]{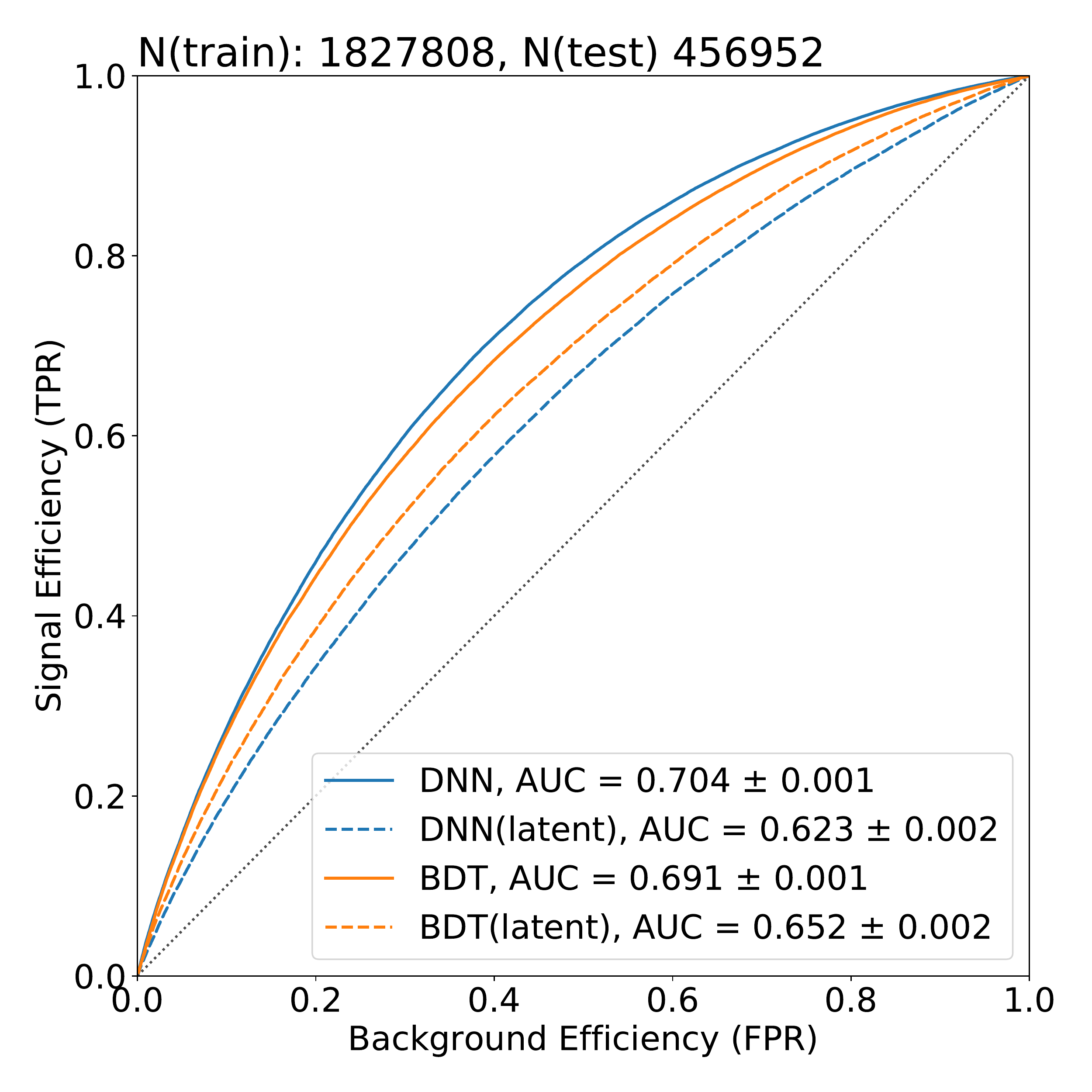}
    \caption{The computed ROCs  for the BDT and DNN using the test data set. The test data set was split into 5 subsets and for each one the AUC was computed. The mean and uncertainty of the AUCs is depicted in the legend. The models were trained both using the full set of of input features (67) and the reduced set of features (16), constructed from the latent space of one of the two developed Autoencoders (see Sec. \ref{sec:autoencoder}).} 
    \label{fig:classical_hep_roc}
\end{figure}

We employed a Deep NN and BDTs to assess the performance of realistic HEP approaches to the $t\bar{t}H(b\bar{b})$ classification task. The full set of available simulated Monte Carlo samples\footnote{For more details about the data set and pre-processing see the corresponding Sec. \ref{sec:dataset}.} was split for the training ($80\%$) and testing ($20$\%) of the models.  The achieved performance is presented in Fig. \ref{fig:classical_hep_roc}. Apart from the training using all the input features (67) the models were also trained utilising only reduced set of features (16) of the latent space of one of the developed Autoencoders (see Sec. \ref{sec:autoencoder}). The latter serves as one of the performance benchmarks developed in this study to investigate the performance of our quantum models, discussed in detail in Sec. \ref{sec:res}.
It should be highlighted that the training and test samples used with these models are far larger than the ones used in the quantum models that we have developed.

\section{Quantum Classifier Models}
\label{sec:qclas}
In this section, we  introduce the two quantum classifier models that we have used: quantum Support Vector Machines and Variational Quantum Circuits.
Both systems encode the classical input as states in a Hilbert space --- the quantum feature space --- the dimension of which increases exponentially with the number of qubits. This is done, in both cases, using \emph{feature maps} \cite{Schuld2019}: quantum circuits that depend on the input data.
While this encoding is conceptually equivalent in both approaches, the two models differ in the way the quantum state is handled, once the data has been encoded.

\subsection{Quantum Support Vector Machine}

The QSVM is the quantum counterpart of a kernel{\color{blue}-}based SVM classifier~\cite{svmVapnik}, with the fundamental difference that the feature map and, hence, the corresponding kernel, is implemented via a quantum circuit~\cite{Havlicek2019}. The goal is to design  the quantum circuit in such  a way that, firstly, it transforms the input data in  a manner that is exponentially hard to simulate classically; and, secondly, the quantum feature map allows the background and signal events to be more easily distinguishable in the feature space than in the input one. 

\begin{figure}[h]
    \centering
    \includegraphics[width=.85\textwidth]{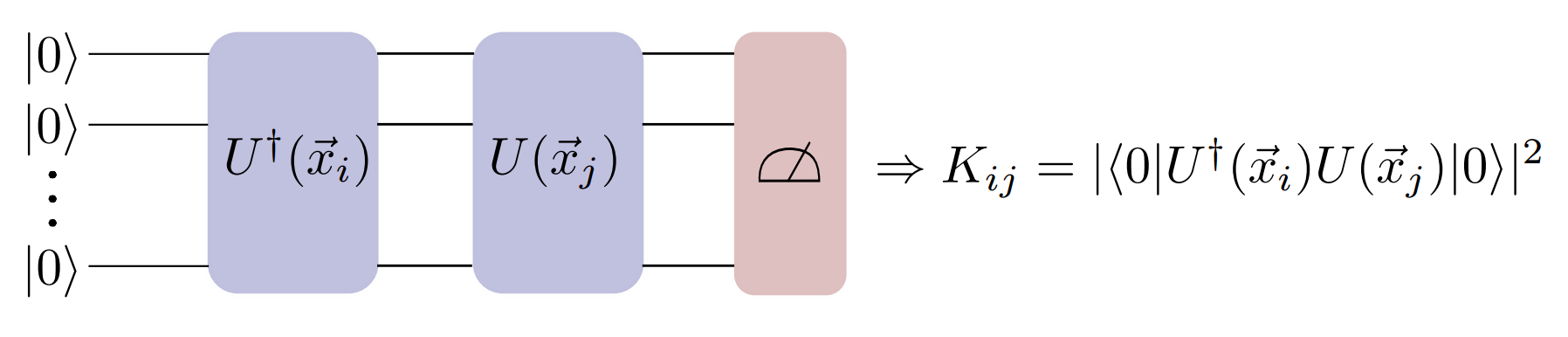}
    \caption{Quantum Circuit for the QSVM, where $\vec{x}_i$ and $\vec{x}_j$ are feature vectors of a given pair of data points, $i$ and $j$. The circuit constructs the kernel matrix elements, $K_{ij}$, by sampling the probability of measuring $|0\rangle \coloneqq |0\rangle^{\otimes\, n^\text{qubits}}$, making the kernel (inner product) quantum.}
    \label{fig:qsvm}
\end{figure}

The loss function of a SVM depends on the inner product of the feature vectors. Specifically, the Lagrangian dual form of the SVM is:
\begin{align}
&\text{maximize}\,\, L(c_1 \ldots c_n) =  \sum_{i=1}^n c_i - \frac 1 2 \sum_{i=1}^n\sum_{j=1}^n y_ic_i(\vec{x}_i\cdot \vec{x}_j)y_jc_j,\\
&\text{subject to } \sum_{i=1}^n c_iy_i = 0,\,\text{and } 0 \leq c_i \leq \frac{1}{2n\lambda}\equiv C\;\text{for all } i.\label{eq:svm}
\end{align}
where $c_i$ are the independent variables of the loss,  $\vec{x}_i$ and $\vec{x}_j$ are feature vectors of a given pair of data points, $i$ and $j$ and $y_i,y_j$ their corresponding labels (e.g. $y_i=1$ for a signal event $i$ and $y_j=-1$ for a background event $j$). 
In this formulation of the SVM loss, the dependence on the inner product between pairs of data points is apparent. Using the kernel trick substitution~\cite{kernel}  $(\vec{x_i}\cdot\vec{x}_j)\to k(\vec{x}_i, \vec{x}_j)\equiv \phi(\vec{x}_i)\cdot \phi(\vec{x}_j)$, where $\phi(\bullet)$ is a feature map, we can directly replace the SVM Kernel with a quantum one that is implemented via a quantum circuit as depicted in Fig. \ref{fig:qsvm}. $C$, or equivalently $\lambda$, is the regularization parameter that tunes the trade-off between mis-classification and width of the SVM margin. Tuning this parameter makes the model robust against over-fitting.

\begin{figure}[thb]
    \centering
    \includegraphics[width=0.95\textwidth]{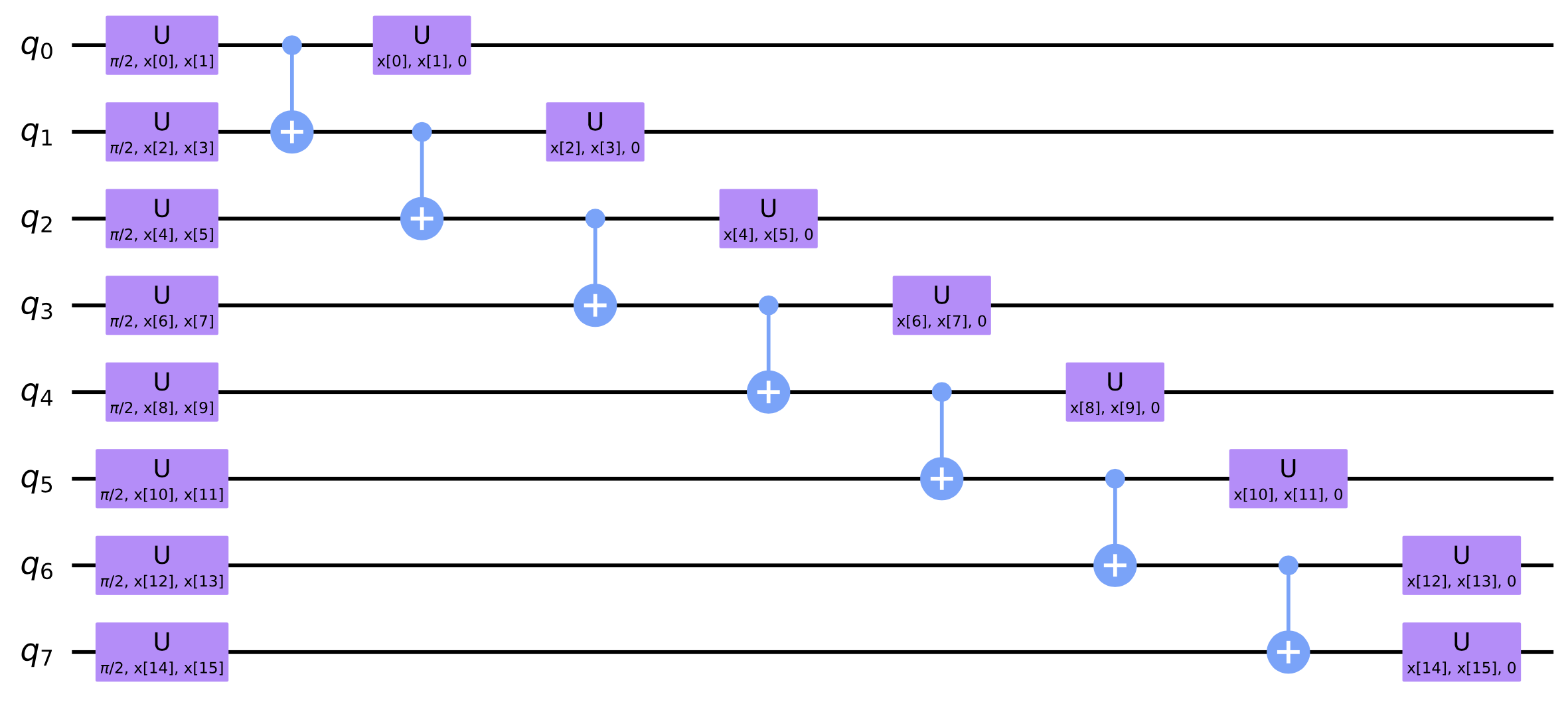}
    \caption{Data encoding circuit serving as feature map for the 8-qubit QSVM implementation. The circuit includes generic unitary 1-qubit  gates that depend on the elements (16) of the data features vector $\vec{x}$, and 2-qubit Control-X (CNOT) gates accomplishing entanglement.}
    \label{fig:u2_reuploading}
\end{figure}

In the QSVM model, the kernel matrix is computed using a quantum computer and the minimization of the SVM loss function is performed on a classical computer. As feature maps, we considered two data embedding circuits. Namely, an amplitude encoding circuit, as described in \cite{Schuld2020cc}, for a 4-qubit and 6-qubit architecture, and the circuit depicted in Fig.~\ref{fig:u2_reuploading} for an 8-qubit architecture. An important aspect of amplitude encoding is its ability to  \emph{exponentially} reduce the required number of qubits: with $N$ qubits, it is able to encode $2^N$ features. Thus, the 4-qubit circuit can encode $2^4=16$ features in the amplitudes of a generic state $\ket{\psi}\in\mathcal{H}^{\otimes 4}$, where $\mathcal{H}$ is the Hilbert space of a single qubit. Similarly the 6-qubit implementation of amplitude encoding is able to embed 64 features to the quantum circuit. In contrast, the designed circuit Fig.~\ref{fig:u2_reuploading} requires 8 qubits but results in a much shallower architecture. This aspect would make the later circuit more suitable of an implementation on a NISQ device.

\subsection{Variational Quantum Circuit}

As initially pointed out in \cite{abbas}, a  VQC can be viewed as a quantum Neural Network. In a VQC, once the feature map has encoded the classical data, the quantum state goes through a circuit with layers of gates that depend on trainable parameters $\vec{\theta}$ and act in a serial manner, mimicking the forward pass of a Neural Network.

\begin{figure}[h]
    \centering
    \includegraphics[width=\textwidth]{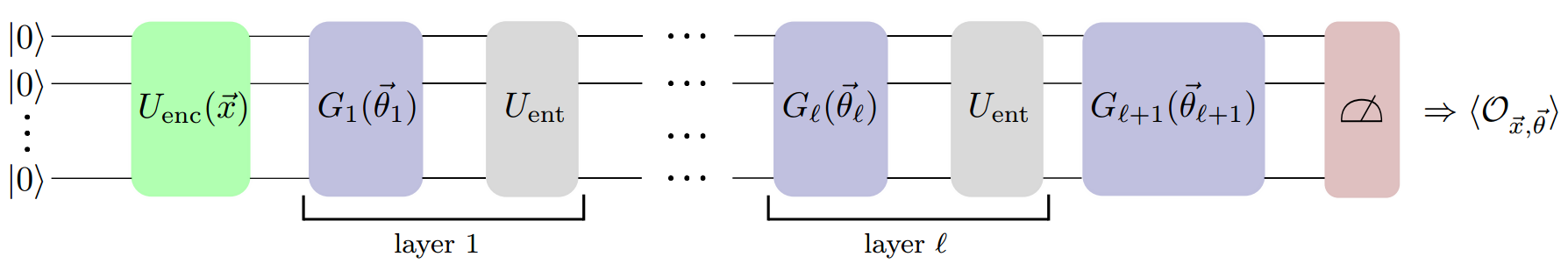}
    \caption{Quantum Circuit for the VQC. $U_\text{enc}$ encodes the data vector $\vec{x}$ into the quantum circuit, then $\ell$ layers of parametrized circuits $G$ and entanglement circuits $U_\text{ent}$ are used. The trainable parameters are $\vec{\theta} = (\vec{\theta}_1,\dots,\vec{\theta}_\ell)$. $\mathcal{O}_{\vec{x},\vec{\theta}}$ is the observable whose expectation value we sample with the quantum device.} 
    \label{fig:vqc}
\end{figure}
The circuit that includes all the operations that depend on the trainable parameters --- and not on the input data --- is called the \emph{variational form}, and it contains a sequence of layers. Each layer $i$ is defined by a sub-circuit $G_i(\vec{\theta}_i)$ that depends on its own trainable parameters $\vec{\theta}_i$ and an entanglement sub-circuit $U_\text{enc}$.
After the variational form, the measurement of a fixed observable $\mathcal{O}$ is performed: its expectation value is the output of the VQC and is used to classify the input.
This is portrayed and summarised in Fig.~\ref{fig:vqc}.

As in the QSVM case, the optimisation of the trainable parameters is classical and consists in the mimimisation of a loss function. Nevertheless, purely quantum optimization approaches have been proposed~\cite{Grad2019} and applied to HEP examples~\cite{blance2020quantum}.

\emph{Data re-uploading},  a novel technique for the development of quantum classifiers, was introduced in~\cite{perez2020data}. Under this approach, instead of using a single instance of the feature map and the variational form, the quantum classification circuit is made up of several repetitions of the traditional VQC scheme --- each with its own classical inputs in the feature map and  trainable parameters.
This allows for a reduction in the number of used qubits, but leads to deeper circuits.

\begin{figure}[h]
    \centering
    \includegraphics[width=0.85\textwidth]{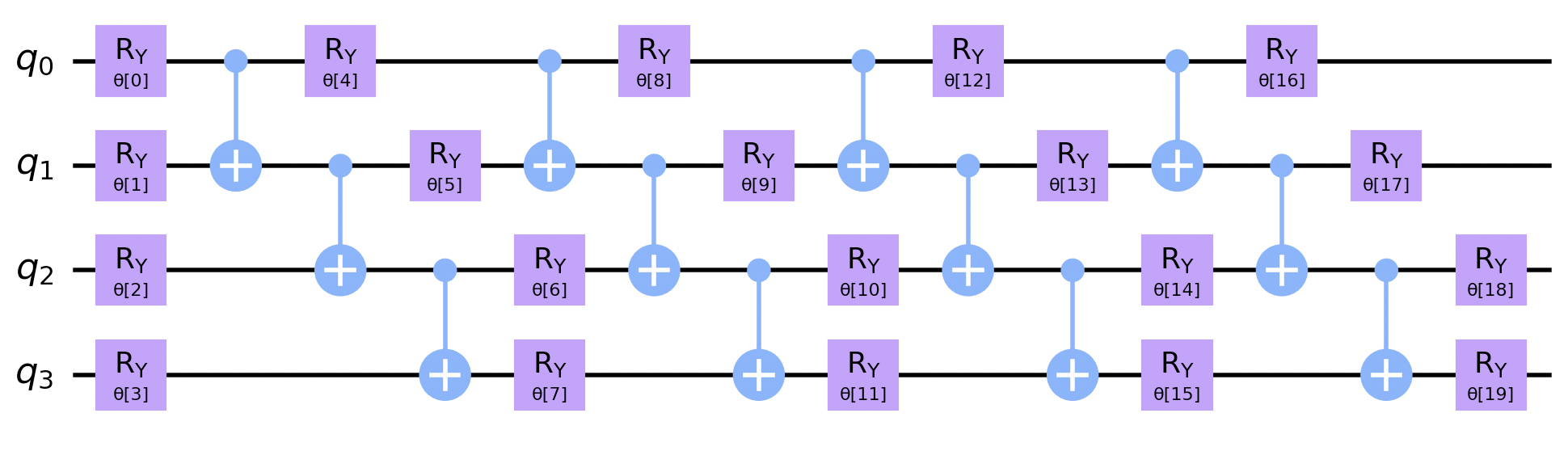}
    \caption{Variational form used in the VQC implementation, dependent on the parameters $\vec{\theta}$.} 
    \label{fig:2local}
\end{figure}

\begin{figure}[h]
    \centering
    \includegraphics[width=1\textwidth]{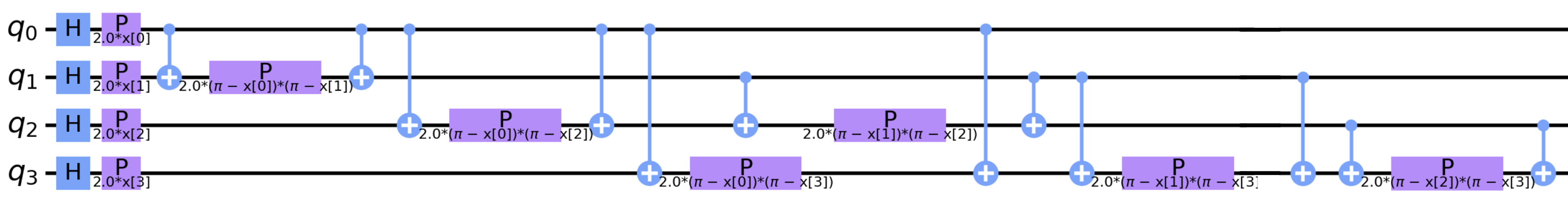}
    \caption{Scheme for the feature map used in the VQC implementation, dependent on data feature vectors $\vec{x}$.} 
    \label{fig:zzfm}
\end{figure}

We have implemented our VQC in a 4-qubit circuit. The variational form that we have used performs $Y$-rotations on each of the qubits by a value specified through a trainable parameter (this would be the $G_i(\vec{\theta}_i)$) alternating with a linear cascade of Controlled-NOT operations (the $U_\text{enc}$) depicted in Fig.~\ref{fig:2local}.
The feature map uses two repetitions of a scheme that combines Pauli gates to encode the data and C-NOT gates to create entanglement. This scheme is represented in Fig.~\ref{fig:zzfm}.

We have designed the VQC to use $8$ input variables. In order to accommodate all of them in this $4$-qubit setup, we have resorted to the data re-uploading technique: using the feature map to load the first four variables followed by an instance of the variational form, and, immediately thereafter, using the same scheme to load the next four variables with an independent set of trainable parameters in the variational form.

Lastly, we measure the first qubit on the computational basis (z). Any state with higher probability of $\ket{0}$ in the first qubit is labelled as background and any state with higher probability of $\ket{1}$, as signal. The chosen loss, is the binary cross-entropy. Since all our experiments were run on an ideal simulator, we have direct access to the real probabilities; in an experimental setup, however, these would have to be computed empirically by running the circuit for a sufficiently large number of shots.

\section{Results}
\label{sec:res}
We used a different training dataset for each of the two methods, but all the tests were carried out on a common collection of five datasets, each with 720 samples. In addition to the two quantum models, whose results are presented in the next two subsections, we also trained some simple classical ML models (SVMs, Logistic Regression, Decision Trees, Random Forests, Multilayer Perceptrons, AdaBoost, kNN, Naive Bayes and QDA) and evaluated them on those same test sets. The training sets used for the classical models were the ones used for the quantum methods, using both the full set of features and a reduced set, 16 or 8 as in the quantum case. In order to have a fair comparison and highlight the effect of the feature extraction step, we have surveyed different techniques to reduce the number of features:  we used the ones obtained by the two AEs, of course, then  the 16 and 8 most informative ones according to their AUC values, and, finally, we applied  some common feature extraction techniques: PCA, KMeans, Truncated SVD, Isomap and Locally Linear Embedding. We also performed grid search to optimize the hyperparameters. Due to a lack of space, we will only show here the results obtained by the best models.

\subsection{Quantum Support Vector Machine}
All developed QSVM models were trained on 576 samples and the regularisation parameter $\lambda$ (see Eq. \ref{eq:svm}) was chosen to be $0.2$ by means of a grid search. The 4-qubit and 8-qubit implementations utilised the 16 latent space features of the PyTorch AE, as described in Sec. \ref{sec:autoencoder}, while a 6-qubit QSVM with amplitude encoding was also developed. The latter is trained on a subset of 64 features out of the original 67; in this case, the 3 individually least informative features ($AUC\sim 0.5$) were discarded. 

\begin{figure}[ht]
\begin{subfigure}{.48 \textwidth}
    \centering
    \includegraphics[width=\textwidth]{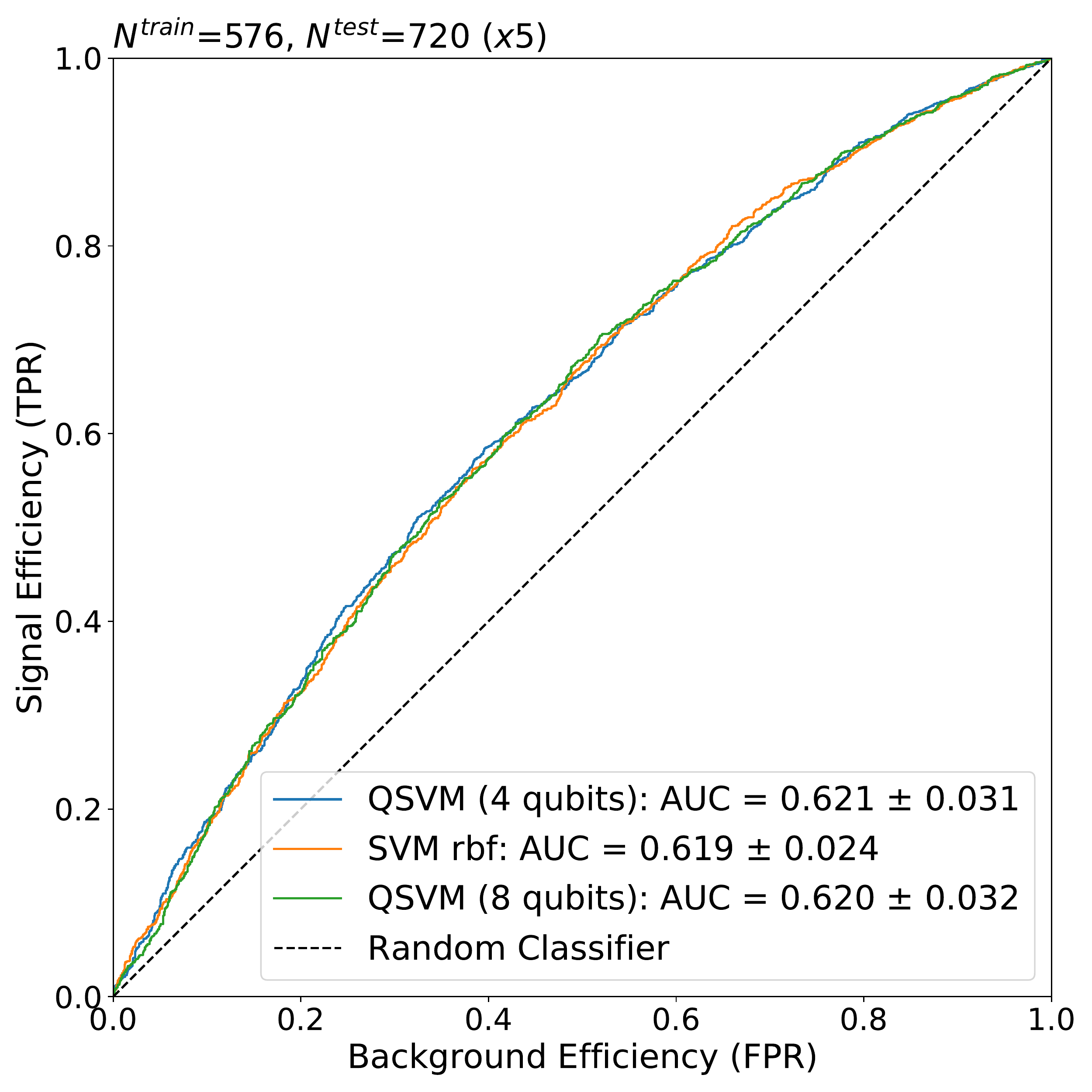}
    \caption{Models trained on the AE latent space features (16).}
    \label{fig:roc_latent}
\end{subfigure}
\begin{subfigure}{.48\textwidth}
    \centering
    \includegraphics[width=\textwidth]{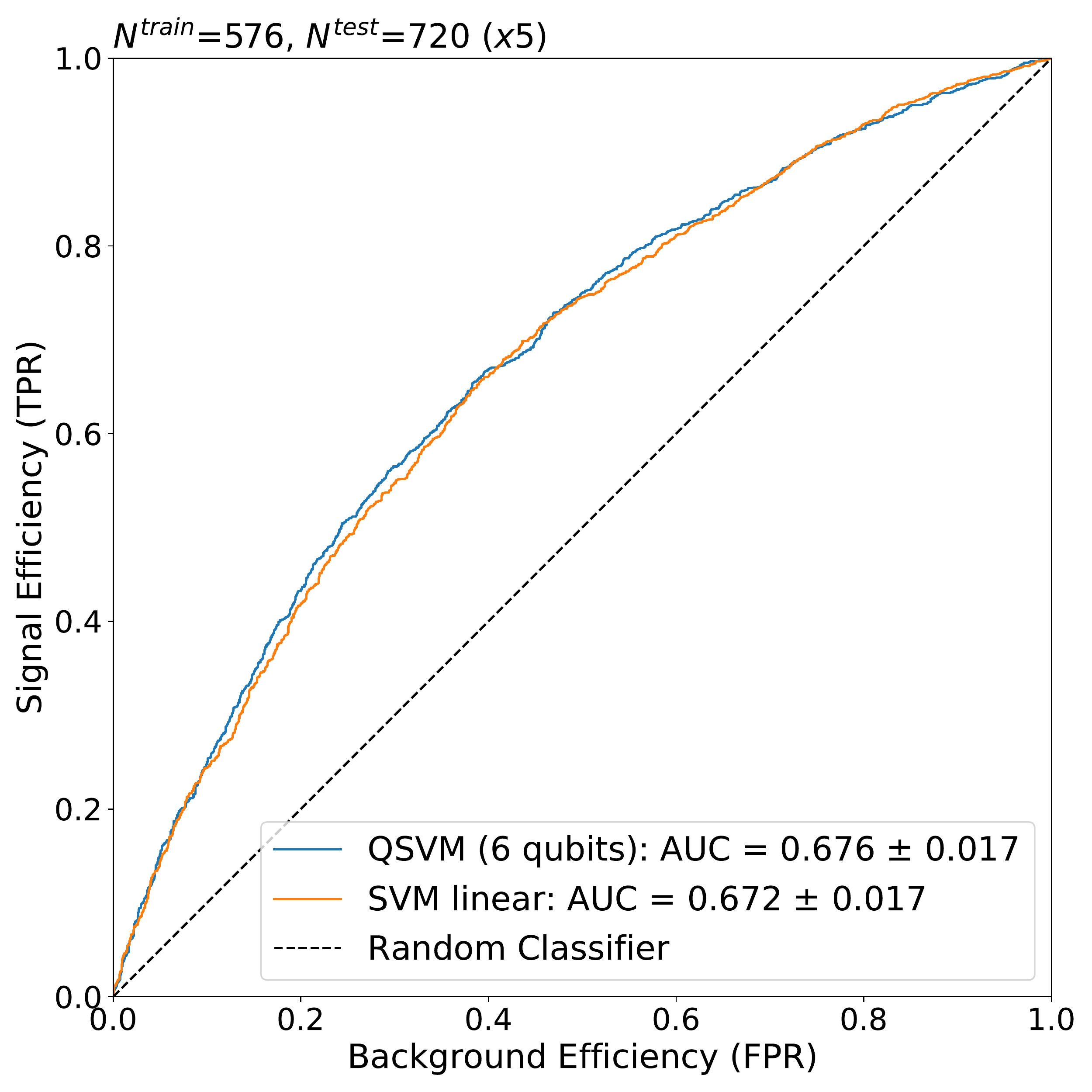}
    \caption{Models trained on the original input features (67), discarding the 3 least informative ones (64).}
    \label{fig:roc_input}
\end{subfigure}
\begin{subfigure}{\textwidth}
    \centering
    \includegraphics[width=.45\textwidth]{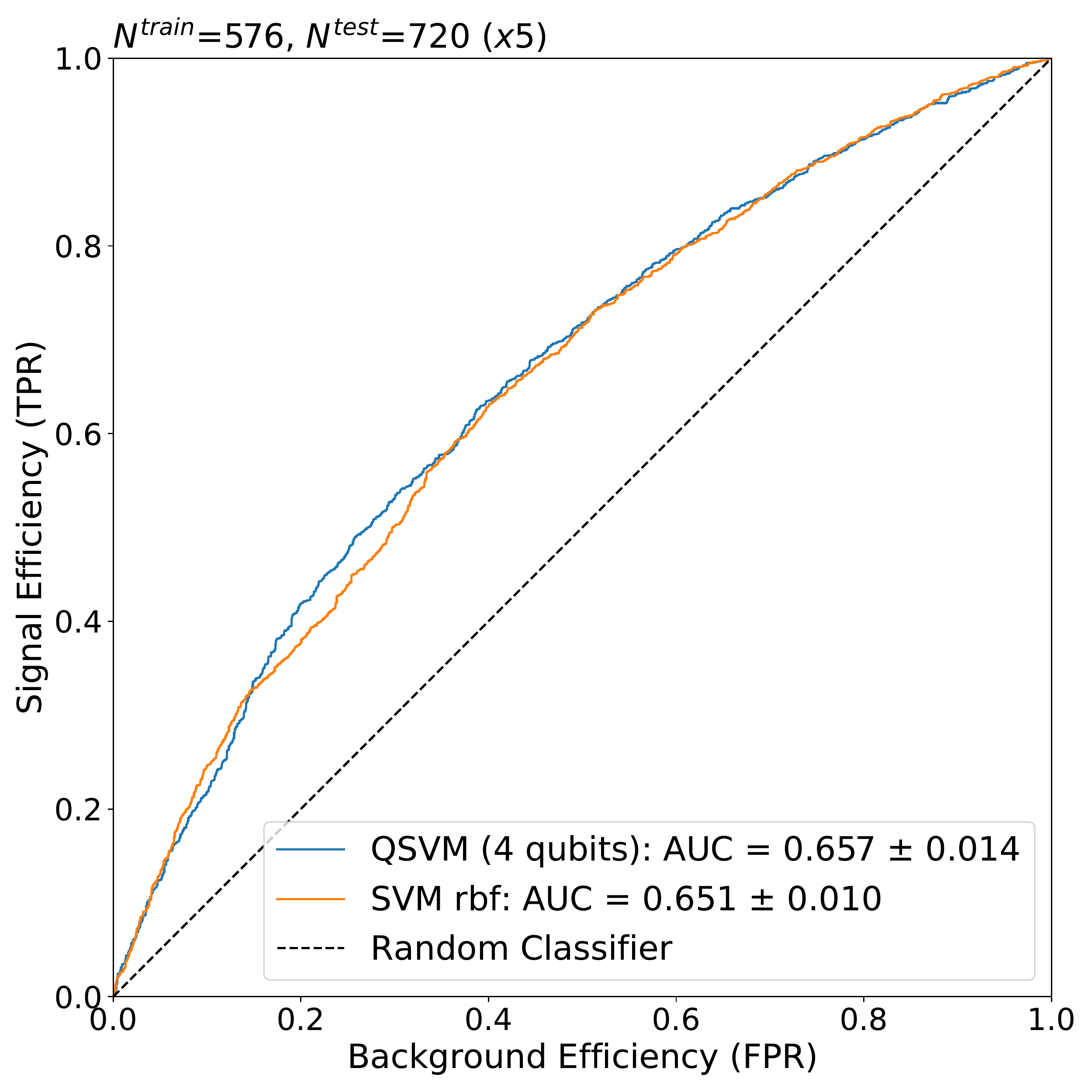}
    \caption{Models trained on 16 selected features of the input space according to their individual AUC values.}
    \label{fig:roc_16input}
\end{subfigure}
\caption{ROC plots for the QSVM models and the corresponding best performing classical models.}
\label{fig:qsvm_roc}
\end{figure}

In Fig.~\ref{fig:qsvm_roc}, we present the performance of the QSVM models with respect to the ROC and AUC values. The ROC is computed using the concatenated test set ($5\times720$ samples) while the mean AUC and its uncertainty are computed across the 5 test data sets. The best performing classical models are also depicted for comparison. In \ref{fig:roc_latent}, the models are trained using the 16 AE latent space features. Both the 4 and 8-qubit models present a similar performance to a classical SVM with a radial base function (rbf) kernel. In ~\ref{fig:roc_input}, the performance of the models trained on 64 of the original 67 features is depicted. Similarly, the 6-qubit QSVM with amplitude encoding shows the same performances as an SVM with a linear kernel. Lastly, \ref{fig:roc_16input} includes the models trained utilising 16 out of the 67 input features. The selection of the 16 features is based on their individual AUC values, which is interpreted as their individual discrimination power for the classification task.

Summarising, the 4-qubit and 8-qubit architectures of the QSVM model behave similarly. In principle, for a NISQ device implementation, one could use the best-suited architecture of the two for the quantum hardware. We observed that the 6-qubits encoding leads to the best performance out of all the QSVM models we developed; this was to be expected, since this architecture uses most of the input information. The 16-best features selected according their individual AUC yield better results than the feature reduction made with the Auto-Encoder, using the 4-qubit QSVM model.

\begin{table}[h]
    \begin{subtable}{.5\textwidth}

\centering
\begin{tabular}{|c|c|}
\hline {\small Feature selection + Model} & AUC \\ \hline\hline 
AUC + QSVM & $0.66\pm0.01$ \\ \hline
PyTorch AE + QSVM & $0.62\pm0.03$ \\ \hline
{\small AUC + SVM rbf} & $0.65\pm0.01$ \\
{\small PyTorch AE + SVM rbf} & $0.62\pm0.02$ \\
{\small KMeans + SVM rbf} & $0.61\pm0.02$ \\ \hline
\end{tabular}
\caption{16 input variables}
\end{subtable}
\begin{subtable}{.5\textwidth}
\centering
\begin{tabular}{|c|c|}
\hline {\small Feature selection + Model} & AUC \\ \hline\hline 
AUC + QSVM & $0.68\pm0.02$ \\ \hline
{\small AUC + Linear SVM} & $0.67\pm0.02$ \\
{\small Logistic Regression} & $0.68\pm0.02$ \\ \hline
\end{tabular}
\caption{64 (QSVM, LSVM) and 67 (LR) input variables}
\end{subtable}
    \caption{Comparison of the AUC value for the QSVM and the best classical models with different feature reduction methods trained on $576$ samples and with an identical or similar number of input variables. SVM rbf stands for ``SVM with a radial base function kernel''.}
    \label{tab:my_label}
\end{table}

\subsection{Variational Quantum Classifier}

In the (classical) optimisation of the variational form parameters, we used the Adam optimiser on $3000$ samples for $70$ epochs with a learning rate of $5\times10^{-3}$ and a batch size of $50$.
Our initial test uses the  eight features of the TensorFlow AE (described in section \ref{sec:autoencoder}) and encodes them in 4 qubits following the {\it data re-uploading} procedure. The resulting AUC was $0.566\pm0.025$, a relatively low value confirming what, already, the analysis of the QSVM performance had suggested: the dimensionality reduction and feature extraction step is not optimal and degrades the quality of the information fed to the VQC. This hypothesis is, indeed, verified by training the same VQC circuit using as input the 8 variables with the highest discrimination power, as determined by their AUC, among the original 67. The corresponding ROC curve is shown in Fig.~\ref{fig:vqc-roc} and compared to the best classical model we trained on the same input (a Random Forest). The corresponding VQC AUC is $0.662\pm0.015$ and represents our best result among the VQC tests we have run. It should be noted, however, that these results are preliminary and the VQC architecture together with the training process need further optimization, which is currently on-going.

\bigskip

{\noindent\begin{minipage}{\textwidth}
\begin{minipage}{0.45\textwidth}
\includegraphics[width=\textwidth]{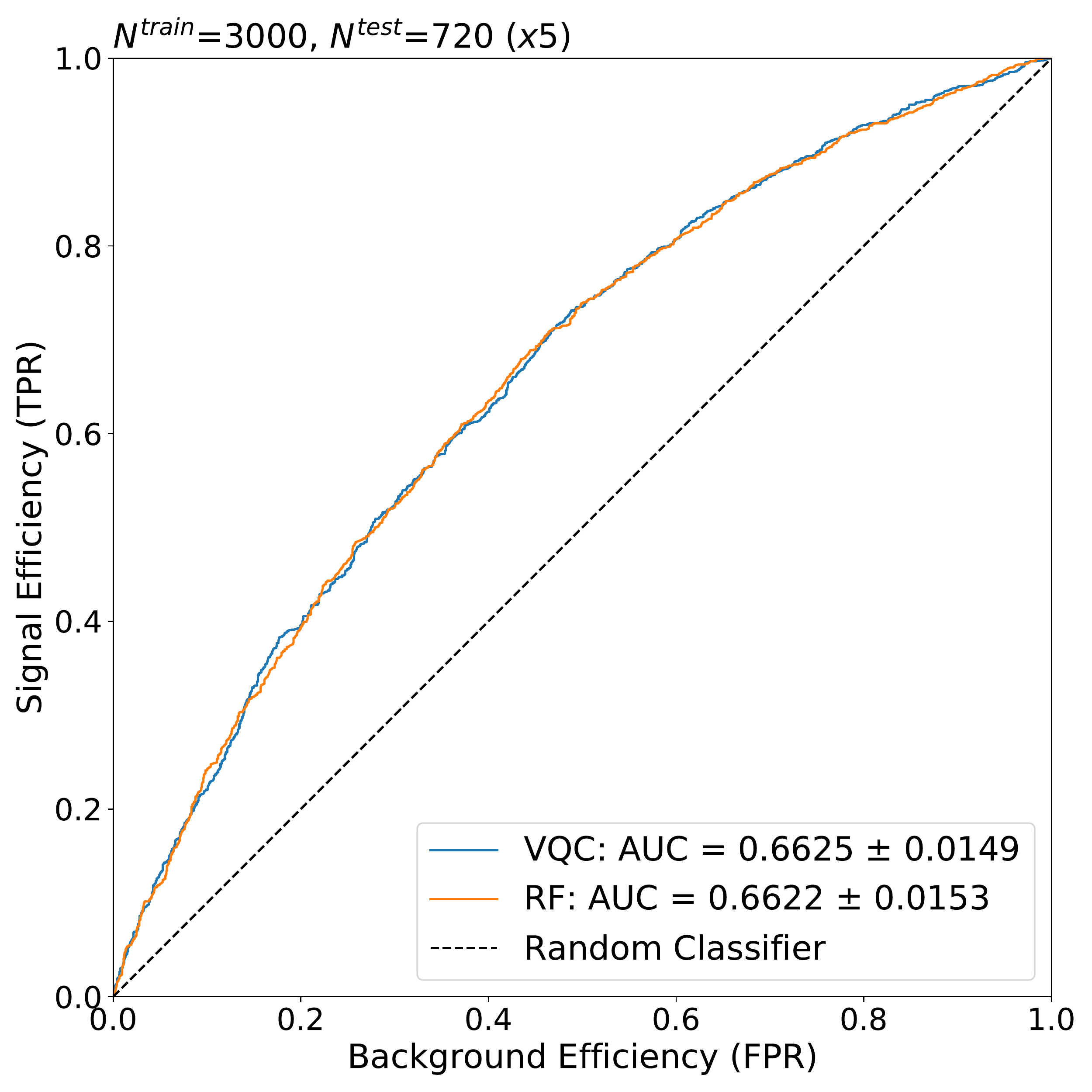}
\captionof{figure}{ROC plot for the VQC and the classical models trained on the 8 variables with more discrimination power (as determined by the AUC).}
\label{fig:vqc-roc}
\end{minipage}
\hfill
\begin{minipage}{0.5\textwidth}
\centering
\begin{tabular}{|c|c|}
\hline {\small Feature selection + Model} & AUC \\ \hline\hline 
AUC + VQC & $0.66\pm0.01$ \\ \hline
{\small AUC + Random Forest} & $0.66\pm0.02$ \\
{\small KMeans + Log. Regr.} & $0.64\pm0.01$ \\
{\small TensorFlow AE + AdaBoost} & $0.63\pm0.03$ \\ \hline
\end{tabular}
\captionof{table}{Comparison of the AUC value of the best VQC with the best classical models using different feature reduction methods with eight input variables trained on $3000$ samples.}
\end{minipage}
\end{minipage}}

\section{Conclusions}
\label{sec:conc}
We have presented results of an initial investigation aimed at understanding strengths and flaws of a quantum approach to the classification of a complex final state of the $t\bar{t}H(b\bar{b})$ process. We explored different quantum algorithms, in terms of classifier architectures (QSVM and VQC), data encoding structures (amplitude encoding, direct encoding and data re-uploading), and data dimensionality reduction strategies. A detailed comparison to several classical approaches, including SVMs, BDTs and Random Forests, shows that, despite the limited number of qubits constrained by capabilities of the NISQ hardware, the quantum classifiers achieve results that are similar, if not slightly better, than classical models trained on the same datasets, in agreement with previous studies~\cite{terashi2021,wu2020application,blance2020quantum,mott2017}.

At the same time, our findings clearly point to the importance of the initial feature extraction step: this is a critical process in classical analysis and it seems to be the same in the (\emph{current}) quantum configuration. In particular, our current AE-based strategy needs further improvement in order to  achieve a better low-dimensionality description of the information contained in the original data set.

In future studies, we will systematically investigate the feature maps and variational forms of the quantum models to maximise performance. Moreover, we expect to simultaneously optimise the developed algorithms for implementations on NISQ devices and assess the effect of hardware noise on model performance.

\emph{Note added}. At the time of submission we became aware of a related work on quantum support vector machines for Higgs analysis~\cite{wu2021}.
\bigskip

\bibliography{references}

\end{document}